\documentclass[12pt]{article}
\usepackage{amsfonts}
\usepackage{graphicx}
\usepackage{mathrsfs}
     \newtheorem{theorem}{Theorem}[section]
     \newtheorem{lemma}{Lemma}[section]

     \newtheorem{definition}{Definition}[section]
     \newtheorem{remark}{Remark}[section]

\begin{document}
\title{\bf Action Minimizing Solutions of The One-Dimensional $N$-Body Problem With Equal Masses \thanks{Supported partially by NSF of China}}

\author{\small\sc Xiang Yu\footnote{Email:xiang.zhiy@gmail.com} \small{and}
\small\sc Shiqing Zhang\footnote{Email:zhangshiqing@msn.com} \\
 \small \it Department of
Mathematics, Sichuan University,
 \small\it Chengdu 610064,  China}

\date{}
\maketitle \vspace*{-9mm}
\begin{center}
\begin{minipage}{5.8in}
\noindent\textbf{Abstract.} When we use variational methods to study the Newtonian $N$-body problem, the main problem is how to avoid collisions.  C.Marchal got a remarkable result, that is, a path minimizing the Lagrangian action functional between two given configurations is always a true (collision-free) solution, so long as the dimension $d$ of physical space $\mathbb{R}^d$ satisfies $d\geq2$. But Marchal's idea can't apply to the case of the one-dimensional physical space. In this paper, we will study the fixed-ends problem for the one-dimensional Newtonian $N$-body problem with equal masses to supplement Marchal's result. More precisely, we first get the isolated property of collision moments for a path minimizing the action functional between two given configurations, then, if the particles at two endpoints have the same order, the path minimizing the action functional is always a true (collision-free) solution; otherwise, although there must be collisions for any path, we can prove that there are at most $N! - 1$ collisions for any action minimizing path.
\end{minipage}
\end{center}

\section{Introduction and Main Results}
\indent\par \setcounter{equation}{0}

In  Euclidean space ${\mathbb{R}}^d$, we consider $N \geq 2$ particles with positive masses
, affected by their  gravitational interactions. The equation of motion of the $N$-body problem is written as
\begin{equation}\label{eq:Newton's equation1}
m_k \ddot{q}_k =\sum_{1 \leq j \leq N, j \neq k} \frac{m_jm_k(q_j-q_k)}{|q_j-q_k|^{3}}.
\end{equation}
where $m_k$ is the mass and $q_k$
the position of the $k$-th body. Since these equations
are invariant by translation, we can assume that the center of masses is at the origin. Firstly, we set some notations and describe preliminary results that will be
needed later. Let $\mathcal{X}_d$ denote the space of configurations for $N$ point particles in Euclidean space $\mathbb{R}^d$ with dimension $d$, whose center of masses is at the origin, that is, $\mathcal{X}_d = \{ q = (q_1,\cdots, q_N)\in (\mathbb{R}^d)^N: \sum_{k = 1}^{N} {m_k q_k} = 0  \}$. For each pair of indices $j, k \in\{ 1,\ldots,N\}$, let $\Delta_{(j,k)} $ denote the collision set of the j-th and k-th particles $\Delta_{(j,k)} = \{ q \in \mathcal{X}_d: q_j = q_k \}$. Let $\Delta_d = \bigcup_{j,k} \Delta_{(j,k)}$ be the collision set in $\mathcal{X}_d$. The space of collision-free configurations $\mathcal{X}_d \backslash \Delta_d$ is denoted by $\hat{\mathcal{X}}_d$. Let $\mathbb{T}$ denote the time interval  $ [T_1, T_2] $.By the path space $\Lambda$, we mean the Sobolev space $\Lambda  = H^1(\mathbb{T}, \mathcal{X}_d)$; we denote by $\Lambda (q_i,q_f)$ the space of paths $q(t) \in \Lambda$ beginning in the configuration ${q_i}$ at the moment $T_1$  and ending in the configuration ${q_f}$ at the moment $T_2$. For a motion $q(t)$ of the $N$-body problem, we say there is  a collision at time $t_0$
if, for at least two indices, say $j$ and $k$, $q_k(t) \rightarrow c_k$, $q_l(t) \rightarrow c_l$  as $t \rightarrow t_0$, and  $c_j = c_k$. We now `cluster' the particles according to their limit points, that is, according to which particles are colliding each other. So, let the different limit points be $c_1, \cdots, c_n$, and let $S_k=\{j \in \{1, \cdots, N\}: q_j(t) \rightarrow c_k ~{as}~ t \rightarrow t_0\},~ k=1, \cdots, n$. We consider the opposite of the potential energy (force function) defined by
\begin{equation}
U(q) = \sum_{k<j} {\frac{m_k m_j }{|q_k - q_j|}}.
\end{equation}
The kinetic energy is defined (on the tangent bundle of $\mathcal{X}_d$) by $K = \sum_{j=1}^{N} {\frac{1}{2}{m_j |\dot{q}_j|^2}}$, the total energy is $E = K- U$ and the Lagrangian is $L(q,\dot{q}) = L = K + U = \sum_j \frac{1}{2} m_j |\dot{q}|^2  + \sum_{k<j}{\frac{m_k m_j}{|q_k - q_j|}}$.
Given the Lagrangian L, the positive definite functional $\mathcal{{A}}:\Lambda \rightarrow \mathbb{R} \cup \{+\infty\}$
defined by
\begin{equation}
\mathcal{{A}}(q) = \int_{\mathbb{T}}{ L(q(t),\dot{q}(t)) dt}
\end{equation}
 is termed as action functional (or the Lagrangian action).

The action functional $\mathcal{{A}}$ is of class $C^1$ on the collision-free space $\hat{\Lambda} (q_i,q_f) \subset \Lambda (q_i,q_f) $. Hence the critical point of $\mathcal{{A}}$ in $\hat{\Lambda} (q_i,q_f) $ is a classical solution (of class $C^2$) of  Newtonian equations
\begin{equation}\label{eq:Newton's equation}
m_j \ddot{q}_j = \frac{\partial U}{\partial q_j}.
\end{equation}

From the viewpoint of the  Least Action Principle, action minimizing solutions of the N-body problem are the most important and the simplest, so it is natural to search for minimizers of the Lagrangian
action joining two given configurations in a fixed time.
It's worth noticing that a lot of results have been founded by the action minimization methods just in recent years, please see \cite{barutello2004action,chen2001action,chen2003binary,chen2008existence,chenciner2002action,Chenciner2002,Chenciner2003,
chenciner2000remarkable,chenciner2000minima,ferrario2004existence,long2000geometric,maderna2009globally,marchal2002method,yu2013saari,zhang2002variational,zhang2004nonplanar}
and the references therein. Recently, the interest in this problem
has grown considerably due to the discovery of the figure eight
solution \cite{chenciner2000remarkable}.

Since the potential of the $N$-body
problem is singular at collision configurations, the main problem involved in variational minimizations is that collision could occur for an action minimizer, even if the set of collision times has necessarily  zero measure, the
system undergoes a collision of two or more bodies, which prevents it form being
a true solution.  Some techniques are created to overcome the difficulty,  ultimately, one got a major advance (essentially due to  Christian Marchal) in this subject. More specifically, the advance is
 the following remarkable theorem \cite{marchal2002method,Chenciner2003,ferrario2004existence}.

\begin{theorem}\label{Marchal}{\emph{(Marchal)}}
Given the initial moment $T_1$,the final moment $T_2$ $(T_2>T_1)$ and two corresponding N-body configurations $q_i = (q_{i1}, \cdots, q_{iN})$, $q_{f} = (q_{f1}, \cdots, q_{fN})$  in $\mathbb{R}^d$ $(d > 1)$, an action minimizing path joining
$q_{i}$ to $q_{f}$
in time $ T_2 - T_1 $ is collision-free for $t \in (T_1, T_2)$.
\end{theorem}
This theorem, together with the lower semicontinuity
of the action, implies in particular that there always exists a collision-free minimizing solution joining two given collision-free N-body configurations in
a given time.

The idea of Christian Marchal is to compare the average of the Lagrangian action for local deformations in all possible directions for a local isolated collision with the original Lagrangian action. Roughly speaking, Marchal's idea is as following : let $a={2}$, by $\frac{1}{2\pi}\int^{2\pi}_0{(\frac{1}{|a+e^{\sqrt{-1}\theta}|}-\frac{1}{|a|})}<0$(i.e., the average of the Lagrangian action on local deformations is smaller than the original Lagrangian action), then there must be some $\theta$ satisfying ${\frac{1}{|a+e^{\sqrt{-1}\theta}|}<\frac{1}{|a|}}$; however, in the case of $d=1$, we  have $\frac{\frac{1}{|a+1|}+\frac{1}{|a-1|}}{2}-\frac{1}{|a|}>0$(i.e., the average of Lagrangian action on local deformations is bigger than the original Lagrangian action), so Marchal's idea can't apply to the case of the one-dimensional
physical space. In fact, Marchal's method is local, but the fixed-ends problem for the one-dimensional Newtonian $N$-body problem is a more global problem, since given two collinear configurations, if the particles at two configurations have different order, then any path joining two given  configurations suffers collisions for topological reasons, hence Marchal's theorem does not hold  for the one-dimensional
physical space. Fortunately, the one-dimensional Newtonian $N$-body problem has its particular characteristics, in particular, the fact that all collinear central configurations are non-degenerate gives us the other facility. Thus, in this paper, by using a different approach, we will study the fixed-ends (Bolza) problem for the one-dimensional Newtonian $N$-body problem. More precisely, we will prove that the path  minimizing the Lagrangian action functional between two given configurations is always a true (collision-free) solution of the one-dimensional $N$-body problem, if the particles at two endpoints have the same order, where, we say that the particles at configurations $q_i = (q_{i1}, \cdots, q_{iN})$ and $q_{f} = (q_{f1}, \cdots, q_{fN})$ have \textbf{the same order} if $q_{ij}-q_{ik} \geq 0 \Leftrightarrow q_{fj}-q_{fk} \geq 0$ for any $j \neq k$, in other words, the relations $q_{ij} > q_{ik}$ and$q_{fj} < q_{fk}$ can't hold for any $j \neq k$ at the same time.  In particular, if $q_{j_1} < q_{j_2} < \cdots < q_{j_N}$, we call $({j_1} , {j_2} , \cdots , {j_N})$ is \textbf{the order of the configuration} $(q_1, q_2, \cdots, q_N)$. This requirement is necessary, since it is obvious that there must be collisions for any path if the particles at two endpoints have different order.

In this paper, we will study the fixed-ends problem for the one-dimensional Newtonian $N$-body problem with equal masses. Our main results are the following Propositions.

\begin{theorem}\label{collisionform}
Suppose the critical path $q(t)$ of the Lagrangian action for the one-dimensional Newtonian $N$-body problem has a collision at some moment $t_0$,  every corresponding colliding cluster $S_k$ has $n_k$ elements. If the collision is isolated at time $t_0$ for some right neighborhood or left neighborhood of $t_0$, then we have the following results for some right neighborhood or left neighborhood of $t_0$:\\
if $n_k=1$, that is,the cluster $S_k$ is singleton, the body in the cluster is not in a collision, let $j \in S_k$, then $q_j(t) = q_j(t_0) + \dot{q}_j(t_0)(t - t_0) + o(t - t_0)$;\\
if $n_k \geq 2$, let $j \in S_k$, then $q_j(t) = q_j(t_0) + s_j(t - t_0)^\frac{2}{3} + o((t - t_0)^\frac{2}{3})$, where $s_j, j \in S_k $ is a central configuration for the particles corresponding to the colliding cluster $S_k$.
\end{theorem}
\begin{remark}
Our results depend strongly on the fact that all collinear central
configurations are non-degenerate.
\end{remark}

\begin{theorem}\label{collisionisolated}
Suppose the action minimizer $q(t)$ of the Lagrangian action for the one-dimensional Newtonian $N$-body problem  with equal masses has a collision at moment $t_0$, then the collision moment $t_0$ is isolated, that is, there exists some $\varepsilon > 0$, $q(t)$ is collision-free in $(t_0 - \varepsilon, t_0 + \varepsilon)$ except at time $t_0$. Hence there are at most finitely many collision moments for the fixed-ends (Bolza) problem.
\end{theorem}

\begin{remark}
There are some studies about the isolated collision for the general $N$-body problem(see \cite{Chenciner2003,ferrario2004existence,venturelli2002application}). However, all the results of them only said that: \textbf{there exists an isolated collision} for the general $N$-body problem. Our results show that we can say more about the one-dimensional Newtonian $N$-body problem with equal masses: all the collisions are isolated and finite.
\end{remark}

\begin{theorem}\label{MarchalR1}
For the one-dimensional  $N$-body problem with equal masses,
given the initial moment $T_1$,the final moment $T_2$ $(T_2>T_1)$ and two corresponding N-body configurations $q_i = (q_{i1}, \cdots, q_{iN})$, $q_{f} = (q_{f1}, \cdots, q_{fN})$  in $\mathbb{R}^1$, if $q_i$, $q_{f}$  have the same order in $\mathbb{R}^1$, then the action minimizing path of the fixed-ends problem  joining
$q_{i}$ to $q_{f}$
in time $ T_2 - T_1 $ is collision-free for $t \in (T_1, T_2)$.
\end{theorem}

\begin{theorem}\label{collisiontimes}
If the given two configurations $q_i$, $q_{f}$  have the different order in $\mathbb{R}^1$, then the action minimizing path of the fixed-ends problem with equal masses joining
$q_{i}$ to $q_{f}$
in time $ T_2 - T_1 $ has some collisions for some $t \in (T_1, T_2)$, but there are at most $N! - 1$ collision moments in $(T_1, T_2)$.
\end{theorem}

\begin{remark}
Our results and methods remain valid for more general force function defined by
$
U(q) = \sum_{k<j} {\frac{m_k m_j }{|q_k - q_j|^\alpha}}
$, where $\alpha$ is any  positive real number such that $0<\alpha<2$ .
\end{remark}

It is natural to ask
the following  questions.

{\bf Question.} 1. Do the \textbf{Theorem \ref{collisionisolated},\ref{MarchalR1} and \ref{collisiontimes}} hold for the one-dimensional $N$-body problem with any masses? 2.Given two configurations which have the different order in $\mathbb{R}^1$ and a time $T = T_2 - T_1 > 0$, what is the largest number of collision times in $(T_1, T_2)$? Is the largest number of collision times in $(T_1, T_2)$ one? The similar questions can be asked for the fixed-ends problem with any masses.

We hope that the answers of these questions are all positive.

The paper is structured as follows. \textbf{Section 2} introduces some  definitions and some lemmas, \textbf{Section 3} gives the proofs of the main results  by using the concepts and results introduced in \textbf{Section 1} and \textbf{Section 2}.

\section{Some Definitions and Some Lemmas}
\indent\par
\setcounter{equation}{0}

%Throughout this paper, we assume that $T\in \mathbb{N}$ is a fixed and finite time horizon and trading only takes place at times $k=0,1,...,T$. Let $(\Omega,\mathcal{F},p)$ be a probability space and a $\sigma$-field $\mathcal{F}_t$ be the available information at time $t$.

In this section, we give some definitions and  recall some classical results.

The first one is  the important concept of the central configuration \cite{wintner1941analytical},
\begin{definition}
A configuration $q=(q_1,\cdots,q_N)\in {\mathcal{X}}_d\setminus\Delta_d$ is called a central configuration if there exists a constant $\lambda\in {\mathbb{R}}$ such that
\begin{equation}\label{centralconfiguration}
\sum_{j=1,j\neq k}^N \frac{m_jm_k}{|q_j-q_k|^3}(q_j-q_k)=-\lambda m_kq_k,1\leq k\leq N,
\end{equation}

the value of $\lambda$ in (\ref{centralconfiguration}) is uniquely determined by
\begin{equation}
\lambda=\frac{U(q)}{I(q)},
\end{equation}
where
\begin{equation}
I(q)=\sum_{1\leq j\leq N} m_j|q_j|^2.
\end{equation}
\end{definition}

Let us recall that, for a motion $q(t)$ of $N$-body problem, we say there is a collision at time $t_0$
if as $t \rightarrow t_0$, $q_j(t) \rightarrow c_j, ~ j \in \{1, \cdots, N\}$ and for at least two different indices, say $j$ and $k$ such that $c_j = c_k$. Without loss of generality, we can assume that the time $t$ approach $t_0$ from the right of $t_0$, that is, we think $t \rightarrow {t_0}+$. Denote the different limit points by $c_1, \cdots, c_n$, and classify the indices according to particles colliding each other,let $S_k=\{j \in \{1, \cdots, N\}: q_j(t) \rightarrow c_k ~{as}~ t \rightarrow {t_0}+\}$, and assume $S_k$ has $n_k $  elements for  $k=1, \cdots, n$; then we say that every $S_k$ is a \textbf{colliding cluster} of particles. Let $M_k=\sum_{j\in S_k} m_j$ be the total mass of particles in cluster $S_k$ and $\bar{c}_k=\sum_{j\in S_k} m_jq_j/M_k$  be the center of mass of the particles in $S_k$.

When $S_k$ has $n_k\geq 2$  elements, if $j \in S_k$, let $r_{(k)j}(t) = \frac{q_j - c_k}{(t-t_0)^\frac{2}{3}}$, then we call  $r_{(k)}(t)=(r_{(k)l_1}(t), \cdots, r_{(k)l_{n_k}}(t))$ be the \textbf{normalized configuration} corresponding to the colliding cluster $S_k$, where  $\{l_1, \cdots, l_{n_k}\} = S_k$.
%And we define the functions $I_k$, $U_k$ and $K_k$ as follows:
%\begin{equation}
%I_k=\sum_{j \in S_k} m_j|q_{j}-\bar{c}_k|^2.
%\end{equation}
%\begin{equation}
%U_k = \sum_{i,j \in S_k, i \neq j} {\frac{m_i m_j }{|q_{i} - q_{j}|}}.
%\end{equation}
%\begin{equation}
%K_k=\sum_{j \in S_k} \frac{1}{2}m_j|\dot{q}_{j}|^2.
%\end{equation}
Let
\begin{equation}
\textbf{CC}_k := \{r_{(k)}: \sum_{j \in S_k,j \neq i} \frac{m_j}{|r_{(k)j}-r_{(k)i}|^3}(r_{(k)j}-r_{(k)i})  = -\frac{2}{9}r_{(k)i}, i \in S_k \}
\end{equation}
be the set of the central configuration corresponding to colliding cluster $S_k$, where we assume the value of $\lambda$ which only affects the size of the central configuration to be $\frac{2}{9}$, note that the center of mass of $r_{(k)}$ is zero.

Before giving the proofs of the main results of this paper, some lemmas are needed. we recall some classical results concerning a motion $q(t)$ of $N$-body problem in some neighborhood of isolated collision instant $t_0$.

The first one says that all collision orbits of $N$-body problem in some neighborhood of isolated collision instant $t_0$ have the  property that $r_{(k)}(t) \rightarrow \textbf{CC}_k$ as $t \rightarrow t_0$, where $r_{(k)}(t)$ and $\textbf{CC}_k$ are respectively the normalized configuration of the collision orbit and the set of the central configuration corresponding to colliding cluster $S_k$.
\begin{lemma}\label{collisionconfiguration}
Suppose a colliding cluster $S_k$ have $n_k \geq 2$ elements, let $r_j(t) = \frac{q_j - c_k}{(t-t_0)^\frac{2}{3}}$ for any $j \in S_k$, be the \textbf{normalized configuration}. Then for every converging sequence $r(t_j)=(r_{l_1}(t_j), \cdots, r_{l_{n_k}}(t_j))$, where  $l_1, \cdots, l_{n_k}  \in S_k$, $t_j$  belong to some neighborhood of $t_0$ $(j \in \mathbb{N})$, the limit $\lim_{j \rightarrow \infty} r(t_j) := s$
 is a central configuration.
\end{lemma}
\begin{remark}
This result is classical(see \cite{saari1984manifold,ferrario2004existence} for a proof). Because of the called ($Painlev\acute{e}$-$Wintner$) infinite spin problem(see \cite{wintner1941analytical,saari1981manifolds,saari1984manifold,chenciner1998collisions,Chenciner2003}et al), in general, one can not get a better result.
\end{remark}

The second one states the special property, which we need, of the one-dimensional Newtonian $N$-body problem.
\begin{lemma}[\cite{saari1980role}]\label{collinearequilateralnon-degenerate}
All collinear central
configurations are non-degenerate in $\mathbb{R}^d$.
\end{lemma}

Then, in the following, we get the important result which says that,  for a isolated collision of particles, not only does $r_{(k)}(t) \rightarrow \textbf{CC}_k$ as $t \rightarrow t_0$, but also there is a central configuration $s \in \textbf{CC}_k$ so that $r_{(k)}(t) \rightarrow s$ as $t \rightarrow t_0$, so long as all  central
configurations are non-degenerate.

\begin{lemma}\label{collisionconfiguration}
For the one-dimensional  $N$-body problem, suppose a colliding cluster $S_k$ have $n_k \geq 2$ elements, let $r_j(t) = \frac{q_j - c_k}{(t-t_0)^\frac{2}{3}}$ for any $j \in S_k$, be the {normalized configuration}. Then $\lim_{t \rightarrow t_0}r(t)$ exists, the limit $s := \lim_{t \rightarrow t_0}r(t)$
 is a central configuration, furthermore, $s$ and $r(t)$ have the same order.
\end{lemma}

{\bf Proof of Lemma \ref{collisionconfiguration}:}

It's similar to a particular case of the results of Saari \cite{saari1984manifold}, we can get {\bf lemma \ref{collisionconfiguration}} by using the unstable  manifold  theorem  for  a normally  hyperbolic  invariant set (Hirsch  et al. \cite{hirsch1970invariant}) and {\bf Lemma \ref{collinearequilateralnon-degenerate}}.\\
$~~~~~~~~~~~~~~~~~~~~~~~~~~~~~~~~~~~~~~~~~~~~~~~~~~~~~~~~~~~~~~~~~~~~~~~~~~~~~~~~~~~~~~~~~~~~~~~~~~~~~~~~~~~~~~~~\Box$

\begin{remark}
There are some methods to study this important problem(see \cite{wintner1941analytical,siegel1971lectures,sperling1969real,saari1981manifolds,saari1984manifold,ElBialy1990Collision,chenciner1998collisions,Chenciner2003,ferrario2004existence}et al). To our knowledge, \textbf{Lemma \ref{collisionconfiguration}} was not definitely stated. Since all collinear central
configurations are non-degenerate, we apply the idea of D.Saari (the unstable  manifold  theorem  for  a normally  hyperbolic  invariant set) to simply get  the result.
\end{remark}

The last lemma is about the existence of isolated collisions for the general $N$-body problem.
\begin{lemma}[\cite{Chenciner2003,ferrario2004existence}]\label{existenceisolatedcollisions}
Suppose the action minimizer $q(t)$ of the Newtonian $N$-body problem has collisions in a time interval, then there must exist an isolated collision in this time interval.
\end{lemma}

Using above lemmas, we will give the proofs of our main results in the next section.

\section{The Proofs of Main Results}
\setcounter{equation}{0}
 \indent\par
In this section, we give the proofs of main results in this paper.

{\bf Proof of Theorem \ref{collisionform}:}

This result easily comes from \textbf{Lemma \ref{collisionconfiguration}}.\\
$~~~~~~~~~~~~~~~~~~~~~~~~~~~~~~~~~~~~~~~~~~~~~~~~~~~~~~~~~~~~~~~~~~~~~~~~~~~~~~~~~~~~~~~~~~~~~~~~~~~~~~~~~~~~~~~~\Box$

First of all, let's establish a lemma to simplify the proofs of other theorems.
\begin{lemma}\label{collisionlesslemma}
Given the initial moment $T_1$,the final moment $T_2$ $(T_2>T_1)$ and two corresponding N-body configurations $q_i = (q_{i1}, \cdots, q_{iN})$, $q_{f} = (q_{f1}, \cdots, q_{fN})\in \mathcal{X}_1 \backslash \Delta_1$ which have the same order in $\mathbb{R}^1$. Suppose a path $q(t) \in \Lambda (q_i,q_f)$ has only one collision moment $t_0$ in $(T_1, T_2)$, then the path $q(t)$ cannot be an action minimizing path of the fixed-ends problem  joining
$q_{i}$ to $q_{f}$
in time $ T_2 - T_1 $.
\end{lemma}
{\bf Proof of Lemma \ref{collisionlesslemma}:}

By using reduction to absurdity, assume that the path $q(t)$ is an action minimizing path of the fixed-ends problem  joining
$q_{i}$ to $q_{f}$
in time $ T_2 - T_1 $. Without loss of generality, we can assume  that $q_1(t)< q_2(t) < \cdots < q_N(t)$ for $t \in [T_1, T_2]\backslash \{t_0\}$ and $q_1(t_0) \leq q_2(t_0) \leq \cdots \leq q_N(t_0)$.

Let $x_k(t) = q_{k+1}(t) - q_{k}(t)$ for $k \in  \{1 , \cdots , N-1 \}$ and $M= m_1 + m_2 + \cdots + m_N $, then  $x(t) = ( x_1(t),  x_2(t), \cdots,  x_{N-1}(t))$ is an action minimizing path of the fixed-ends problem  joining
$x_{i}=x(T_1)$ to $x_{f}=x(T_2)$
in time $ T_2 - T_1 $
for the action functional
\begin{equation}
\mathcal{{F}}(x) = \int_{\mathbb{T}}{ \sum_{1 \leq l<k \leq N}\frac{m_k m_l}{2M} {[{ |\sum_{l \leq j \leq k-1}\dot{x}_j|^2}+ \frac{2M }{|\sum_{l \leq j \leq k-1}{x}_j|}]} dt}
\end{equation}
In fact, by Lagrangian identity, we have
\begin{eqnarray}
\mathcal{{A}}(q) & = & \int_{\mathbb{T}}{ L(q(t),\dot{q}(t)) dt}\nonumber\\
& = &\int_{\mathbb{T}}{ [\frac{1}{2(\sum_{1 \leq j \leq N})m_j} \sum_{1 \leq l<k \leq N} m_k m_l |\dot{q}_k - \dot{q}_l|^2 + \sum_{1 \leq l<k \leq N}\frac{ m_k m_l}{|{q}_k - {q}_l|}] dt}\nonumber\\
& = & \int_{\mathbb{T}}{ \sum_{1 \leq l<k \leq N} \frac{m_k m_l}{2M} {[{ |\sum_{l \leq j \leq k-1}\dot{x}_j|^2}+ \frac{2M }{|\sum_{l \leq j \leq k-1}{x}_j|}]} dt}\nonumber\\
& = & \mathcal{{F}}(x)\nonumber
\end{eqnarray}

In the following, we will construct another path $y(t)$ which satisfies the same  boundary conditions with $x(t)$, but the value of $\mathcal{{F}}(y)$ is smaller than the value of $\mathcal{{F}}(x)$.

Since we can get similar result by using the following method for any $k \geq 1$ such that $x_k(t) \rightarrow 0$ when $t \rightarrow t_0$, for the sake of  convenience, we only consider that $x_1(t) \rightarrow 0$ when $t \rightarrow t_0$. Then we have $x_1(t) = \alpha (t_0 - t)^{\frac{2}{3}}+o((t_0 - t)^{\frac{2}{3}})$ for some left neighborhood of $t_0$ and $x_1(t) = \beta (t - t_0)^{\frac{2}{3}}+o((t - t_0)^{\frac{2}{3}})$ for some right neighborhood of $t_0$ from \textbf{Theorem \ref{collisionform}}, where $\alpha$, $\beta$ are appropriate positive numbers. Let $A = \frac{m_1(M-m_1)}{2M}$ and $B = \sum_{3\leq k \leq N}\frac{m_1m_k}{M}\sum_{2\leq j \leq k-1}\dot{x}_j$, from \textbf{Theorem \ref{collisionform}} we know that
\begin{itemize}
  \item if $x_j(t) \rightarrow 0$ when $t \rightarrow t_1$ for some $j \in \{2, \cdots, N-1\}$, then $B = \frac{d(\tilde{\alpha}(t_0 - t)^{\frac{2}{3}}+o((t_0 - t)^{\frac{2}{3}}))}{dt}$ for some left neighborhood of $t_0$ and $B = \frac{d(\tilde{\beta}(t - t_0)^{\frac{2}{3}}+o((t - t_0)^{\frac{2}{3}}))}{dt}$ for some right neighborhood of $t_0$, where $\tilde{\alpha}, \tilde{\beta}$ are appropriate positive numbers;
  \item if $x_j(t) > 0$ for some neighborhood of $t_0$ and any $j \in \{2, \cdots, N-1\}$, then $B = \frac{d(a+b(t - t_0) +o(|t_1 - t|))}{dt}$ for some neighborhood of $t_0$, where $a>0, b$ are appropriate real numbers.
\end{itemize}
Then it is easy to know that the inequality
\begin{equation}\label{ercishi}
A\dot{x}^2_1+ B\dot{x}_1 > 0
\end{equation}holds in some neighborhood of $t_0$. For sufficiently small positive number $\delta$, there are two sufficiently small positive numbers $\epsilon, \varepsilon $ such that $x_1(t_1-\epsilon)=x_1(t_1+\varepsilon)= \delta$, ${x}_1(t) \leq \delta$ for $t \in [t_1-\epsilon, t_1+\varepsilon]$ and the interval $ [t_1-\epsilon, t_1+\varepsilon]$  is in this neighborhood of $t_0$ for the {inequality (\ref{ercishi})} holds.
Furthermore, we have the inequalities
\begin{equation}
\frac{1 }{|x_1 +\sum_{2 \leq j \leq k-1}{x}_j|} \geq \frac{1 }{|\delta+\sum_{2 \leq j \leq k-1}{x}_j|}
\end{equation} for
$t \in [t_1-\epsilon, t_1+\varepsilon]$ and any $3\leq k \leq N$.

Let $y_1(t)=\delta$ for $t \in [t_1-\epsilon, t_1+\varepsilon]$, $y_1(t)= x_1(t)$ for $t \in [T_1, T_2] \backslash [t_1-\epsilon, t_1+\varepsilon]$, and $y_j(t)= x_j(t)$ for $t \in [T_1, T_2]$ and $2 \leq j \leq N-1$. Let $y(t) = ( y_1(t),  y_2(t), \cdots,  y_{N-1}(t))$, then we know
\begin{eqnarray}
\mathcal{{F}}(x) - \mathcal{{F}}(y) & = & \int^{t_1+\varepsilon}_{t_1-\epsilon}{ \sum_{1 \leq l<k \leq N}\frac{m_k m_l}{2M} {[{ |\sum_{l \leq j \leq k-1}\dot{x}_j|^2}+ \frac{2M }{|\sum_{l \leq j \leq k-1}{x}_j|}]} dt}\nonumber\\
& - & \int^{t_1+\varepsilon}_{t_1-\epsilon}{ \sum_{1 \leq l<k \leq N}\frac{m_k m_l}{2M} {[{ |\sum_{l \leq j \leq k-1}\dot{y}_j|^2}+ \frac{2M }{|\sum_{l \leq j \leq k-1}{y}_j|}]} dt}\nonumber\\
& = &\int^{t_1+\varepsilon}_{t_1-\epsilon}{  {[A\dot{x}^2_1+ B\dot{x}_1 +\sum_{3\leq k \leq N}\frac{m_k m_1 }{|x_1+\sum_{2 \leq j \leq k-1}{x}_j|}]} dt}\nonumber\\
& - & \int^{t_1+\varepsilon}_{t_1-\epsilon}{ \sum_{3\leq k \leq N}\frac{m_k m_1 }{|\delta+\sum_{2 \leq j \leq k-1}{x}_j|} dt}\nonumber\\
& > &0\nonumber
\end{eqnarray}
Hence the path $q(t)$ is not an action minimizing path of the fixed-ends problem  joining
$q_{i}$ to $q_{f}$
in time $ T_2 - T_1 $.\\
$~~~~~~~~~~~~~~~~~~~~~~~~~~~~~~~~~~~~~~~~~~~~~~~~~~~~~~~~~~~~~~~~~~~~~~~~~~~~~~~~~~~~~~~~~~~~~~~~~~~~~~~~~~~~~~~~\Box$

Henceforth, we think all the particles have equal mass, i.e., we assume $m_1 = m_2 = \cdots = m_N = m$.

{\bf Proof of Theorem \ref{collisionisolated}:}

By using reduction to absurdity, without loss of generality, let $t_0$ be an instant at which collision times accumulate for some right neighborhood of $t_0$.
By \textbf{Lemma \ref{existenceisolatedcollisions}}, there are infinite isolated collisions in some right neighborhood of $t_0$. Then it's easy to know that
there are three isolated collision moments $t_1$, $t_2$ and $t_3$ ($t_1<t_2<t_3$)such that the collisions at moments $t_1$, $t_2$ and $t_3$ have the same colliding clusters and the same order,
i.e., as $t \rightarrow t_i(i\in \{1,2,3\})$, there exist different limit points $c_{i1}, \cdots, c_{in}$ such that $S_{ik}=\{j \in \{1, \cdots, N\}: q_j(t) \rightarrow c_{ik} ~{as}~ t \rightarrow {t_i}\}$ and $S_{1k}=S_{2k}=S_{3k}$ for  $k=1, \cdots, n$, furthermore, (without loss of generality) $c_{i1}< \cdots< c_{in}$ for $i\in \{1,2,3\}$. Given $k\in \{1, \cdots, n\}$, if the colliding cluster $S_{2k}$ has $n_k\geq 2$ elements, suppose the order of the particles in $S_{2k}$ is $({l_1}, \cdots, l_{n_k})$ for some left neighborhood of $t_2$ and $({j_1}, \cdots, j_{n_k})$ for some right neighborhood of $t_2$, that is, $q_{l_1}(t) < \cdots < q_{l_{n_k}}(t)$ for some left neighborhood of $t_2$ and $q_{j_1}(t) < \cdots < q_{j_{n_k}}(t)$ for some right neighborhood of $t_2$, where $\{{l_1}, \cdots, l_{n_k}\} = \{{j_1}, \cdots, j_{n_k}\}=S_{2k}$. If $({l_1}, \cdots, l_{n_k}) \neq ({j_1}, \cdots, j_{n_k})$, assume $\tau_k$ is a permutation from $({l_1}, \cdots, l_{n_k})$ to $({j_1}, \cdots, j_{n_k})$, let
\begin{equation}
(h_{j_1}(t), \cdots, h_{j_{n_k}}(t)) = (q_{\tau_k(l_1)}(t),  \cdots, q_{\tau_k(l_{n_k})}(t))
\end{equation} for $t \in [t_1, t_{2}]$.
If $({l_1}, \cdots, l_{n_k})= ({j_1}, \cdots, j_{n_k})$, or if the colliding cluster $S_{2k}$ has $n_k=1$ element, that is,the cluster $S_{2k}$ is singleton, thus the body in the cluster is not in a collision, the permutation $\tau_k$ can be chosen as unit transformation, then still let
\begin{equation}
(h_{j_1}(t), \cdots, h_{j_{n_k}}(t)) = (q_{\tau_k(l_1)}(t),  \cdots, q_{\tau_k(l_{n_k})}(t))
\end{equation} for $t \in [t_1, t_{2}]$.

Finally, let $h(t)=(h_1(t), \cdots,h_N(t))$ for $t \in [t_1, t_{2}]$ and $h(t)=(q_1(t), \cdots,q_N(t))$ for $t \in [t_2, t_{3}]$, then $h(t)$ is a path in the Sobolev space $H^1([t_1,t_3], \mathcal{X}_1)$ with fixed-ends such that $h(t_1)=q(t_1)$ and $h(t_3)=q(t_3)$. Indeed, by the construction of $h(t)$, the relations $h(t_1)=q(t_1)$ and $h(t_3)=q(t_3)$ are obvious; by the continuity of $h(t)$ at $t=t_2$, it's easy to know that $h(t)$ has weak derivative $\dot{h}(t)$ in $[t_1, t_{3}]$, furthermore, $\dot{h}(t)$  is square integrable in $[t_1, t_{3}]$ by applying the finiteness of the Lagrangian action.

Let us recall that, if all the particles have the same masses,  there is an obvious fact: \textit{suppose $\tau$ is a permutation of $(1, 2, \cdots, N)$, let $r(t)=(r_1(t), r_2(t), \cdots, r_N(t)) = (q_{\tau(1)}(t), q_{\tau(2)}(t), \cdots, q_{\tau(N)}(t))$, if $m_1 = m_2 = \cdots = m_N$, then
\begin{equation}
 \int^{T_2}_{T_1}{ L(q(t),\dot{q}(t)) dt} = \int^{T_2}_{T_1}{ L(r(t),\dot{r}(t)) dt},
\end{equation}}
Since the path $q(t)$ is an action minimizing path, we know that the path $h(t)$ is an action minimizing path in the Sobolev space $H^1([t_1,t_3], \mathcal{X}_1)$ with fixed-ends $h(t_1)=q(t_1)$ and $h(t_3)=q(t_3)$. In particular,  the path $h(t)$ is an action minimizing path in the Sobolev space $H^1([t_2-\epsilon,t_2+\epsilon], \mathcal{X}_1)$ with fixed-ends $h(t_2-\epsilon)$ and $h(t_2+\epsilon)$ for all the sufficiently small $\epsilon>0$. By choosing any sufficiently small $\epsilon>0$, we have a path $h(t)$ such that: the action minimizing path $h(t) \in H^1([t_2-\epsilon,t_2+\epsilon], \mathcal{X}_1)$ has only one collision moment $t_2$ in $(t_2-\epsilon,t_2+\epsilon)$, the fixed-ends $h(t_2-\epsilon)$, $h(t_2+\epsilon)\in \mathcal{X}_1 \backslash \Delta_1$ and have the same order in $\mathbb{R}^1$. However, this contradicts with \textbf{Lemma \ref{collisionlesslemma}}.

In conclusion, if the action minimizing path $q(t)$ of the one-dimensional Newtonian $N$-body problem with equal masses has collisions, then every collision is isolated. Since the set of collision times is closed, we know there are at most finitely many collision moments for the fixed-ends (Bolza) problem.\\
$~~~~~~~~~~~~~~~~~~~~~~~~~~~~~~~~~~~~~~~~~~~~~~~~~~~~~~~~~~~~~~~~~~~~~~~~~~~~~~~~~~~~~~~~~~~~~~~~~~~~~~~~~~~~~~~~\Box$

{\bf Proof of Theorem \ref{MarchalR1}:}

First of all, let's establish a lemma to simplify the proof.

\begin{lemma}\label{lemma3}
Given the initial moment $T_1$,the final moment $T_2$ $(T_2>T_1)$ and two corresponding N-body configurations $q_i = (q_{i1}, \cdots, q_{iN})$, $q_{f} = (q_{f1}, \cdots, q_{fN})$ which have the same order in $\mathbb{R}^1$, suppose the path $q(t) \in \Lambda (q_i,q_f)$ has  collision in $(T_1, T_2)$, and the collision moments  in $(T_1, T_2)$ are respectively $t_1, t_2, \cdots, t_n$ $(T_1 < t_1 < \cdots < t_n < T_2)$. Then there is some path $h(t)\in \Lambda (q_i,q_f)$ such that $\{t_1, \cdots, t_n\}$ are collision moments in $(T_1, T_2)$ and the order of $h(t)$ are the same for all the time $t \in [T_1, T_2]$. Furthermore, if all the particles have the same masses, then
\begin{equation}
 \int^{T_2}_{T_1}{ L(q(t),\dot{q}(t)) dt} = \int^{T_2}_{T_1}{ L(h(t),\dot{h}(t)) dt} .
\end{equation}
\end{lemma}
{\bf Proof of Lemma \ref{lemma3}:}

It's easy to know that, there is some path $g(t)$ which has the same order with $q_i$ and $q_f$ in $\mathbb{R}^1$ and $g(t)$ is collision-free for $t \in (T_1, T_2)$. Suppose the order of the orbit $g(t)$ for $t \in (T_1, T_2)$ is $(j_1, \cdots, j_N)$, that is, $g_{j_1}(t) < \cdots < g_{j_N}(t)$, where $\{j_1, \cdots, j_N\} = \{1, \cdots, N\}$. Without loss of generality, we can assume that $(j_1, j_2, \cdots, j_N)=(1, 2, \cdots, N)$.
Let $t_0 = T_1$ and $t_{n+1} = T_2$, suppose the order of the orbit $q(t)$ for $t \in (t_k, t_{k+1})$ is $(j_{k1}, \cdots, j_{kN})$, that is, $q_{j_{k1}}(t) < \cdots < q_{j_{kN}}(t)$, where $k \in \{0, \cdots, n\}$. Suppose $\tau_k$ is a permutation from $(j_{k1}, \cdots, j_{kN})$ to $(1, 2, \cdots, N)$, let
\begin{equation}
h^{(k)}(t)=(h^{(k)}_1(t), h^{(k)}_2(t), \cdots, h^{(k)}_N(t)) = (q_{\tau_k(1)}(t), q_{\tau_k(2)}(t), \cdots, q_{\tau_k(N)}(t))
\end{equation} for $t \in (t_k, t_{k+1})$.
Firstly, it is easy to know that
\begin{equation}
\lim_{t \rightarrow {t^+_{0}}}h^{(0)}(t)=q_i ,
\lim_{t \rightarrow {t^-_{n+1}}}h^{(n)}(t)=q_f
\end{equation}
In the following, we prove that
\begin{equation}
\lim_{t \rightarrow {t^-_{k+1}}}h^{(k)}_j(t) = \lim_{t \rightarrow {t^+_{k+1}}}h^{(k+1)}_j(t)
\end{equation}
for every $j \in \{1, \cdots, N\}$ and $k \in \{0, \cdots, n-1\}$.

In fact, from $
h^{(k)}_j(t) = q_{\tau_{k}(j)}(t)
$ for $t \in (t_k, t_{k+1})$ and $
h^{(k+1)}_j(t) = q_{\tau_{k+1}(j)}(t)
$ for $t \in (t_{k+1}, t_{k+2})$, it is easy to know that we  only need to prove the relation $q_{\tau_{k}(j)}(t_{k+1}) = q_{\tau_{k+1}(j)}(t_{k+1})$. For the sake of a contradiction, we can suppose that $q_{\tau_{k}(j)}(t_{k+1}) > q_{\tau_{k+1}(j)}(t_{k+1})$ or $q_{\tau_{k}(j)}(t_{k+1}) < q_{\tau_{k+1}(j)}(t_{k+1})$. If $q_{\tau_{k}(j)}(t_{k+1}) > q_{\tau_{k+1}(j)}(t_{k+1})$, from $h^{(k)}_l(t) > h^{(k)}_j(t) $ for $N \geq l > j$,$t \in (t_k, t_{k+1})$,
we have $q_{\tau_{k}(l)}(t_{k+1}) = \lim_{t \rightarrow {t^-_{k+1}}}h^{(k)}_l(t) \geq \lim_{t \rightarrow {t^-_{k+1}}}h^{(k)}_j(t) = q_{\tau_{k}(j)}(t_{k+1}) > q_{\tau_{k+1}(j)}(t_{k+1})$. Hence $h^{(k+1)}_{\tau^{-1}_{k+1}\tau_{k}(l)}(t) = q_{\tau_{k}(l)}(t) > q_{\tau_{k+1}(j)}(t)= h^{(k+1)}_j(t)$ for every $l$ such that $N \geq l \geq j$,$t \in (t_{k+1}, t_{k+1} + \epsilon)$, where $\epsilon$ is some sufficiently small positive number. So we have $\tau^{-1}_{k+1}\tau_{k}(l) > j$ for for every $l$ such that $N \geq l \geq j$, but there are at most $N - j$ number  larger than $j$ in $\{1, 2, \cdots, N\}$, this is a contradiction. If $q_{\tau_{k}(j)}(t_{k+1}) < q_{\tau_{k+1}(j)}(t_{k+1})$, it is similar to get a contradiction. So we have
\begin{equation}
\lim_{t \rightarrow {t^-_{k+1}}}h^{(k)}_j(t) = \lim_{t \rightarrow {t^+_{k+1}}}h^{(k+1)}_j(t)
\end{equation}
for every $j \in \{1, \cdots, N\}$ and $k \in \{0, \cdots, n-1\}$.

Let $h(t)=h^{(k)}(t)
$ for $t \in (t_k, t_{k+1})$,  $h(t_{k})=\lim_{t \rightarrow {t^+_{k}}}h^{(k)}(t)
$ for $1 \leq k \leq n$, $h(T_{1})=\lim_{t \rightarrow {t^+_{0}}}h^{(0)}(t)
$, $h(T_{2})=\lim_{t \rightarrow {t^-_{n+1}}}h^{(n)}(t)
$, then $h(t) \in \Lambda (q_i,q_f)$ and $\{t_1, \cdots, t_n\}$ are collision moments in $(T_1, T_2)$ and the order of $h(t)$ are the same for all the time $t \in [T_1, T_2]$.

Furthermore, since all the particles have the same masses, we have
\begin{equation}
 \int^{T_2}_{T_1}{ L(q(t),\dot{q}(t)) dt} = \int^{T_2}_{T_1}{ L(h(t),\dot{h}(t)) dt} .
\end{equation}

From the above, \textbf{Lemma \ref{lemma3}} holds.\\

In the following, we prove \textbf{Theorem \ref{MarchalR1}} by using \textbf{Lemma \ref{lemma3}}.

By using reduction to absurdity, suppose the action minimizing path $q(t)$ has  collision moments in $(T_1, T_2)$, the collision moments in $(T_1, T_2)$ are respectively $t_1, t_2, \cdots, t_n$ $(T_1 < t_1 < \cdots < t_n < T_2)$. Furthermore, we can assume  that $q_1(t)< q_2(t) < \cdots < q_N(t)$ for $t \in (T_1, T_2)\backslash \{t_1 , \cdots , t_n \}$ and $q_1(t_k) \leq q_2(t_k) \leq \cdots \leq q_N(t_k)$ for $k \in  \{1 , \cdots , n \}$ by using \textbf{Lemma \ref{lemma3}}. Then we can find a path $q(t)\in H^1([t_1-\epsilon,t_1+\epsilon], \mathcal{X}_1)$ which has only one collision moment $t_1$ in $(t_1-\epsilon,t_1+\epsilon)$, and the fixed-ends $q(t_1-\epsilon)$, $q(t_1+\epsilon)\in \mathcal{X}_1 \backslash \Delta_1$ have the same order in $\mathbb{R}^1$, so long as the positive number $\epsilon$ is sufficiently small. However, this contradicts with \textbf{Lemma \ref{collisionlesslemma}}.

So we know that, for the N-body problem with equal masses,  given two moments and corresponding  configurations which have the same order in $\mathbb{R}^1$, the action minimizing path of the fixed-ends problem joining
two configurations
 is collision-free for $t \in (T_1, T_2)$.\\
$~~~~~~~~~~~~~~~~~~~~~~~~~~~~~~~~~~~~~~~~~~~~~~~~~~~~~~~~~~~~~~~~~~~~~~~~~~~~~~~~~~~~~~~~~~~~~~~~~~~~~~~~~~~~~~~~\Box$

{\bf Proof of Theorem \ref{collisiontimes}:}

Suppose the action minimizing orbit $q(t)$ has  collision in $(T_1, T_2)$,
the collision moments in $(T_1, T_2)$ are respectively $t_1, t_2, \cdots, t_n$ $(T_1 < t_1 < \cdots < t_n < T_2)$, let $t_0 = T_1$ and $t_{n+1} = T_2$. Let us investigate  $n+1$ collision-free path sections: $q(t),t \in (t_k, t_{k+1})$, $0 \leq k \leq n$. If  $n>N! - 1$,  then there are two sections which have the same order, suppose the corresponding time intervals are respectively $(t_j, t_{j+1})$ and $ (t_l, t_{l+1})$, $j<l$. Let us  choose two moments $s_1 \in (t_j, t_{j+1})$ and $s_2 \in (t_l, t_{l+1})$, then it is easy to know that the path $q(t)
,t \in [s_1, s_2]$ is an action minimizing orbit of the fixed-ends problem for two moments $s_1, s_2$ and corresponding  configurations $q(s_1), q(s_2)$. However, from  \textbf{Theorem \ref{MarchalR1}}, $q(t)$ is collision-free  in $(s_1, s_2)$, this  contradicts with  $t_{j+1}, t_l \in (s_1, s_2)$.\\
$~~~~~~~~~~~~~~~~~~~~~~~~~~~~~~~~~~~~~~~~~~~~~~~~~~~~~~~~~~~~~~~~~~~~~~~~~~~~~~~~~~~~~~~~~~~~~~~~~~~~~~~~~~~~~~~~\Box$

\section*{Acknowledgements} The authors sincerely thank an anonymous expert for his/her many valuable
comments and suggestions.

%\bibliographystyle{plain}

%\bibliography{C:/Users/YuXiang/Desktop/YuReference}

\end{document}